\documentclass{article}

\usepackage{amstext}
\usepackage{amsmath}
\usepackage{amsthm}
\usepackage{amssymb}
\usepackage{bm}
\usepackage{wrapfig}
\usepackage{amsfonts, amsmath}
\usepackage[utf8]{inputenc}
\usepackage{graphicx}
\usepackage{bbm, comment}
\usepackage{caption}
\usepackage{subcaption}
\usepackage{color}
\usepackage{float}
\usepackage{wrapfig}
\usepackage{lipsum}
\usepackage{enumitem}
\usepackage{url}

\usepackage{thmtools,thm-restate}

\usepackage{hyperref}

\usepackage{tabulary}
\usepackage{booktabs}

\usepackage[top=1.in, bottom=1.in, left=1.in, right=1.in]{geometry}

\usepackage{tikz}
\usepackage{cleveref}
\usepackage{apxproof}

\newcommand{\E}{\mathbb{E}}

\newtheorem*{theorem*}{Theorem}
\newtheorem{theorem}{Theorem}[section]
\newtheorem{lemma}[theorem]{Lemma}
\newtheorem{fact}[theorem]{Fact}
\newtheorem{claim}[theorem]{Claim}

\newtheorem{example}[theorem]{Example}

\newtheorem{definition}[theorem]{Definition}
\newtheorem{observation}[theorem]{Observation}

\newtheoremrep{lemma}[theorem]{Lemma}
\newtheoremrep{claim}[theorem]{Claim}
\newtheoremrep{proposition}[theorem]{Proposition}

\newcommand{\mi}{\mathrm{I}}
\newcommand{\en}{\mathrm{H}}

\title{Multistable Perception, False Consensus, and Information Complements}

\author{Yuqing Kong\\ Peking University}

\date{}

\begin{document}

\maketitle

\begin{abstract}
This paper presents a distributed communication model to investigate multistable perception, where a stimulus gives rise to multiple competing perceptual interpretations. We formalize stable perception as consensus achieved through components exchanging information. Our key finding is that relationships between components influence monostable versus multistable perceptions. When components contain substitute information about the prediction target, stimuli display monostability. With complementary information, multistability arises. We then analyze phenomena like order effects and switching costs. Finally, we provide two additional perspectives. An optimization perspective balances accuracy and communication costs, relating stability to local optima. A Prediction market perspective highlights the strategic behaviors of neural coordination and provides insights into phenomena like rivalry, inhibition, and mental disorders. The two perspectives demonstrate how relationships
among components influence perception costs, and impact competition and coordination behaviors in neural dynamics. 
\end{abstract}

\section{Introduction}

\begin{figure}[!ht]
   \centering
   \begin{subfigure}[t]{0.2\textwidth} 
      \centering
      \includegraphics[height=.5\textwidth]{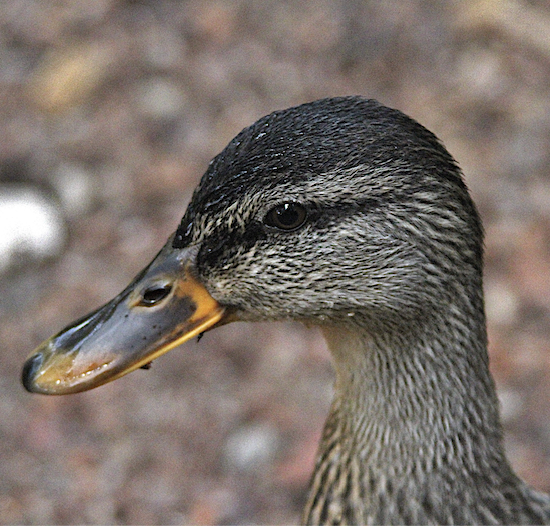}
      \caption{\textbf{Duck} }
      \label{fig:duck}
   \end{subfigure}  
   \hfill 
   \begin{subfigure}[t]{0.2\textwidth}
      \centering
      \includegraphics[height=.5\textwidth]{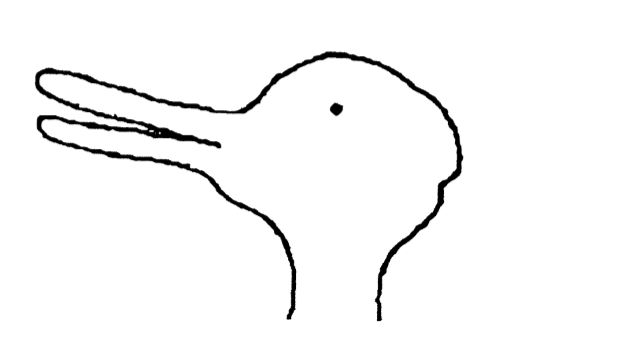}
      \caption{\textbf{Duck-rabbit}}
      \label{fig:dura}
   \end{subfigure}
   \hfill
   \begin{subfigure}[t]{0.2\textwidth} 
      \centering
      \includegraphics[height=.5\textwidth]{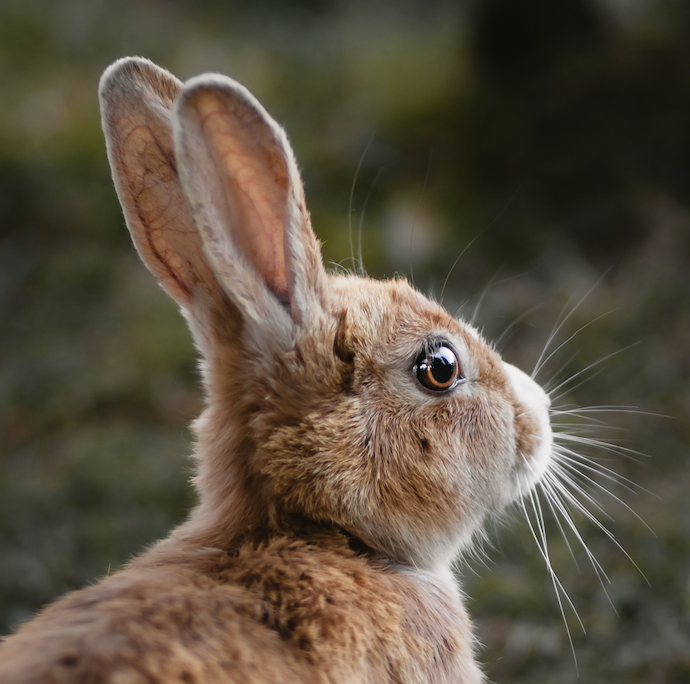}
      \caption{\textbf{Rabbit}  }
      \label{fig:rabbit}
   \end{subfigure} 
   \caption{\textbf{Monostable vs. Multistable}}
   \label{fig:multistable}
\end{figure}
When we look at \Cref{fig:duck} and \Cref{fig:rabbit}, we clearly see a duck and rabbit \footnote{Photo by Frank Eiffert and Tolga Ahmetler on Unsplash.}. However, \Cref{fig:dura} is ambiguous - it can be perceived as either duck or rabbit \cite{kruse2012ambiguity,jastrow1899mind}. This phenomenon is multistable perception - the same stimulus giving rise to multiple interpretations \cite{leopold1999multistable}. 

An important question is: what attributes of a stimulus determine whether it will induce a singular or multiple interpretation? 

One intuitive hypothesis is that simple, straightforward stimuli tend to lead to singular interpretations, while complex, rich stimuli are more likely to lead to multiple views. Complexity can be quantified via entropy \cite{shannon1948mathematical}, which quantifies the richness in a distribution. Hick-Hyman law \cite{hick1952rate,hyman1953stimulus} shows that as the entropy in a choice set increases, response time also increases. However, in our examples, unambiguous figures seem more complex by this metric, as they are drawn from richer distributions. 

Crucially, entropy ignores relationships between components, which can relate in two key ways - substitutes or complements. Substitutes contain redundant information about a target, like survey responses from similar demographics. Complements provide synergistic information about a target, like puzzle pieces. The relationship can be described information-theoretically: the framework of Partial Information Decomposition \cite{wibral2017partial} separates multivariate mutual information into unique, redundant (substitutable), and synergistic (complementary) components. Strategic behaviors in prediction markets illustrate the importance of the two relationships. Traders immediately reveal information when they are substitutes, as the marginal value decreases over time. But they delay releasing when they are complements, as the marginal value increases later \cite{chen2010gaming, chen2016informational}. 

Does the relationship among individual components within stimuli influence the number of stable perceptions? Evidence demonstrates that multistable perceptions arise from interactions between low-level sensory and high-level frontal/parietal regions. Bottom-up explanations propose that perceptual reversals are caused by low-level adaptation or inhibition between competing representations in early visual areas \cite{brown1955rate, kohler1960dynamics,toppino1987selective}. Top-down processes refer to influences from higher cognitive factors that shape perception based on expectations, attention, etc \cite{rock1994ambiguous, leopold1999multistable}. A hybrid model integrates both bottom-up and top-down processes \cite{leopold1999multistable,long2004enduring,sterzer2009neural}. However, these explanatory approaches do not explicitly analyze the role of component relationships in processing ambiguous stimuli.

Understanding this influence requires a model illustrating how the brain integrates separated components. Integrating diverse informational components into a unified whole is considered as a key functionality of consciousness \cite{tononi2016integrated}. While neural network models like convolutional neural networks (CNNs) \cite{lecun1989backpropagation}, recurrent neural networks (RNNs) \cite{elman1990finding}, and long short-term memory networks (LSTMs) \cite{hochreiter1997long} have demonstrated powerful capabilities for information integration, the often ``black box'' nature of their computations makes it difficult to directly analyze the role of component relationships.

To this end, we adopt an abstract distributed communication model. Imagine `detectives' in the brain, resembling neurons, seek to predict events like object category. Each detective accesses distinct information and can share it via a communication protocol, i.e., rules guiding information exchange. In a game-theoretic framework, detectives gain rewards (marginal gain of prediction accuracy) and incur costs to acquire and communicate information. Their behaviors are guided by this incentive structure. As detectives interact, their collective belief\footnote{Notice that the above model does not imply that the neurons really form Bayesian posteriors. Their behaviors can be trained through past experience, \emph{as if} they form Bayesian posteriors. Prior studies showed how plastic synapses are able to estimate posteriors \cite{soltani2010synaptic}.} fluctuates and converges toward consensus through information sharing. 

A key definition is stable perception equals consensus among `detectives'. Prior studies \cite{aumann1976agreeing, geanakoplos1982we,aaronson2005complexity} showed that consensus will be reached through sufficient communication. However, consensus can integrate all or only some information. We define ``regret'' as the amount of unaggregated information in the consensus. Low regret indicates an approximately ``true'' consensus, which is analogous to a complete understanding. High regret represents an incomplete, ``false'' consensus, which is analogous to a superficial comprehension.

With the regret-based characterization, we can answer the original question on monostable versus multistable perceptions. Stimuli eliciting only true consensus will induce singular interpretations, as two true consensuses do not contradict. However, many stimuli likely lead to false consensuses, which are vulnerable to multistability. Prior studies show under substitutes, consensuses are true \cite{kong2023false,frongillo2023agreement}. Thus, substitutes promote monostability\footnote{This aligns with the general idea pervading economics that substitutes lead to uniqueness and stability. With alternative definitions, literature shows complements are connected to equilibrium multiplicity, while substitutes are associated with uniqueness in market \cite{mas1995microeconomic} and cognition game~\cite{pavan2022expectation}.}. In general, we introduce a complexity measure, ``Gestalt complexity''. It is high when the stimulus has high regret stable perceptions. We show that low Gestalt complexity leads to monostable perception.

In addition to the distributed model, predictive coding \cite{rao1999predictive} also offers a Bayesian framework to explain multistable perception through hierarchical prediction-error correction \cite{hohwy2008predictive}. The distributed model provides an alternative theoretic viewpoint for multistable phenomena. In the predictive coding model, stability arises when the sensory data is accurately predicted. In contrast, the distributed approach reveals stable states can integrate information to varying degrees, allowing for stable states representing ``false'' consensus, i.e., superficial understanding. This enables analysis of how interaction patterns between informational components shape singular versus multistable outcomes.

\begin{figure}[!ht]
   \centering
   \begin{subfigure}[t]{0.2\textwidth} 
      \centering
      \includegraphics[height=.5\textwidth]{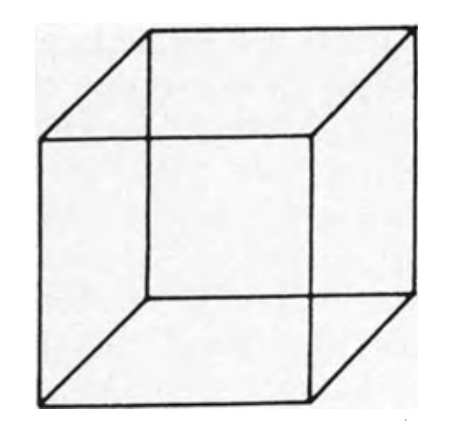}
      \caption{\textbf{The Necker Cube} The Necker cube \cite{boring1942sensation} can be seen as a three-dimensional cube that alternates between appearing closer on the left side and closer on the right side, depending on the viewer's perception. }
      \label{fig:cube}
   \end{subfigure}
   \begin{subfigure}[t]{0.2\textwidth}
      \centering
      \includegraphics[height=.5\textwidth]{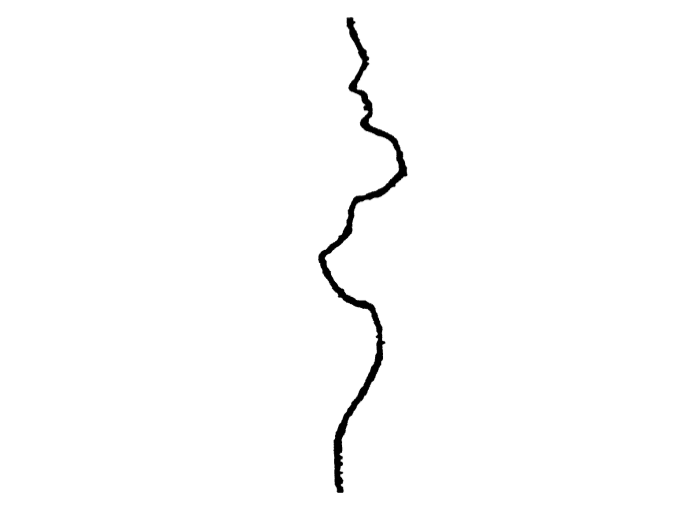}
      \caption{\textbf{Boundary} A line can be perceived as both the rightmost boundary and the leftmost border, depending on the perspective.}
      \label{fig:boundary}
   \end{subfigure}
   \begin{subfigure}[t]{0.2\textwidth}
      \centering
      \includegraphics[height=.5\textwidth]{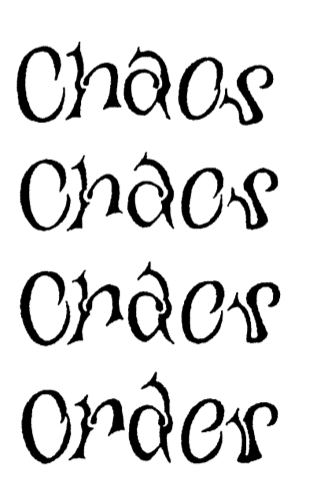}
      \caption{\textbf{Chaos-Order} The picture can be read as "CHAOS, CHAOS, CHAOS" from top to bottom, but when read in the opposite direction, it reads "ORDER, ORDER, ORDER" }
      \label{fig:chaos}
   \end{subfigure}
   \caption{\textbf{Examples of Visual Multistable Perceptions} \cite{kruse2012ambiguity}}
   \label{fig:multistable}
\end{figure}

Multistable perceptions have ``order effects'' - distinct interpretations emerge depending on the sequence of information (\Cref{fig:boundary,fig:chaos}). Order matters not just for external presentation, but also internal memory recall sequence. This relates to judgment biases \cite{kahneman1982judgment}. Prior models illustrate how selected and limited recall of information impacts final conclusions \cite{gennaioli2010comes}. Our communication framework complements this by showing how relationship types shape perception given a particular order. Substitutes lead to the singular interpretation regardless of order. But complements display path dependency - different trajectories of consensus formation lead to distinct interpretations.

We then study the relationship between switching ease and divergence of stable perceptions. We hypothesize more surprising perceptions have higher costs to perceive and erase. We show divergence must be bounded for easy switching between perceptions. The hardest switching is between full and superficial understanding because full understanding is very surprising relative to superficial one.

Finally, we present two additional perspectives. First, the communication protocol can be viewed as an optimization process that aims to balance accuracy and costs, like the information bottleneck framework \cite{tishby2000information}. Stable perception is a local optimum, with substitutes enabling a convex-ish goal where the local optimum is the global optimum. This aligns with work that models multistability as attractors \cite{kelso2012multistability}. 

Balancing value and cost is also the key theme in brain theories in the biological and physical sciences from the free-energy perspective \cite{friston2006free,bullmore2012economy}. Though, unlike prior models, the current model explicitly highlights component relationships. The optimization space is the communication protocols. The cost closely relates to communication complexity - the minimal exchanged bits required for distributed computational processes to jointly solve a problem \cite{yao1979some,kushilevitz1997communication}. The communication cost is important as evidence shows that mental fatigue impairs understanding while preserving access to individual elements~\cite{boksem2008mental}. In real life, we sometimes understand every word of a sentence but do not know the meaning of the sentence. This is because the communication cost to integrate the words is high.  

With an emphasis on component relationships and communication, this optimization-based perspective demonstrates that substitutes produce an easy perception. In contrast, complements lead to false consensus traps where beliefs get stuck far from the truth. Full understanding requires escaping poor local optima, which can be very challenging. 

The second perspective draws an analogy to prediction markets \cite{wolfers2004prediction}, which aggregate distributed information and generate forecasts by leveraging market forces like trading and price signaling. This perspective highlights \emph{strategic} neural coordination using analogy to strategic traders and demonstrates that substitutes produce competitive behaviors of neurons, and complements lead to more collaborative behaviors in neural dynamics. This perspective potentially provides insights into phenomena like rivalry \cite{tong2006neural}, inhibition \cite{aron2007neural}, mental disorders \cite{frith1979consciousness,geschwind2007autism}, and confirmation bias \cite{wason1960failure} by leveraging analyses of the strategic behaviors of traders.

\section{A Distributed Communication Model}
We view the perception process as a distributed communication where multiple agents, resembling neurons, seek to predict a target event $W$ based on the stimulus $\mathbf{x}$ through communication.

\paragraph{Stimulus $\mathbf{X}$, Target $W$, and Information Structure $\theta$} Stimulus $\mathbf{X}$ consists of $n$ signals $X_1, X_2,\cdots, X_n$. $W$ is an event that is relevant to $\mathbf{X}$. $X_1,X_2,\cdots, X_n$ and $W$ are random variables. $\Omega$ denotes $W$'s realization space. For all $i$, $\mathcal{S}_i$ denotes signal $X_i$'s realization space. $\Delta_{\mathcal{S}_1\times \mathcal{S}_2\times\cdots\times \mathcal{S}_n\times\Omega}$ denotes the space of all possible distributions over $X_1,X_2,\cdots, X_n$ and $W$. Information structure $\theta\in \Delta_{\mathcal{S}_1\times \mathcal{S}_2\times\cdots\times \mathcal{S}_n\times\Omega}$ denotes the joint distribution over $X_1, X_2,\cdots, X_n$ and $W$, which describes the relationship among these random variables. $\mathbf{x}=(x_1,x_2,\cdots,x_n)$ denotes a realization of $X_1, X_2,\cdots, X_n$. 

$X_1, X_2, \ldots, X_n$ represent individual sensory components of either an external or internal stimulus. External stimuli include visual images, sounds, text, etc., while internal stimuli include memories. The target variable $W$ refers to perceptual interpretations like object category or sentence meaning.

The information structure $\theta$ characterizes the relationship between components $\mathbf{X}$ and target $W$. It guides belief updates about $W$ as $\mathbf{X}$ are sequentially revealed, and is learned through past experience. $\theta$ captures dependencies and redundancies within $\mathbf{X}$ regarding $W$. Later we will analyze how properties of $\theta$ influence singular versus ambiguous perceptions given an input stimulus $\mathbf{x}$.

\paragraph{The Shared Blackboard Model} Each agent $i$ has access to $X_i$, such that she can acquire possibly partial knowledge of $X_i$. Agents communicate through a \emph{shared blackboard} visible to all iteratively. 
\begin{itemize}
\item At each round $t$, a random agent is picked. She is allowed to write a public message on the board;
\item After the termination of each round, all agents are asked whether they want to continue. If all agents choose to not continue, the protocol stops. 
\end{itemize}

$h^{t}$ denotes the history of all contents written on the blackboard in the first $t$ rounds. 

\paragraph{Communication Protocol} Agent $i$ has a deterministic policy $P_i$ guiding behavior. At round $t$, selected agent $i_t$'s policy $P_{i_t}$ takes the private signal $x_{i_t}$ and the current board contents $h^{t-1}$ as inputs and outputs a message $P_{i_t}(x_{i_t},h^{t-1})$ she will write on the board. The board contents update to $h^t=(P_{i_t}(x_{i_t},h^{t-1}),h^{t-1})$. At round end, $\forall i$, $P_i$ takes $h^t$ and outputs ``stop'' or ``continue''.

The policies $\{P_i\}_i$ and agent picking order $\mathbf{i} = i_1, i_2, \ldots$ constitute communication protocol $\Pi = (\mathbf{i}, \{P_i\}_i)$. Given stimulus $\mathbf{x}$ and protocol $\Pi$, the process is deterministic. We denote by $\Pi(\mathbf{x})$ the final board contents upon termination. Agents know the policies $\{P_i\}_i$ but not the order $\mathbf{i}$.

\begin{example}\label{exa:setting}
We illustrate these concepts with a simple example. There are two agents, Xayla and Yanis, with access to private signals $X$ and $Y$ respectively. The signal spaces are $\mathcal{S}_X = \mathcal{S}_Y = {0,1,2}$. $W$ is a binary event.

The information structure $\theta$ is a joint distribution over $X,Y,W$. Marginally $X,Y \sim \text{Uniform}\{0,1,2\}$. That is, $\Pr[X=x,Y=y]=\frac{1}{9},\forall x,y\in \{0,1,2\}$. For any $x,y$, $\Pr[W=0|X=x,Y=y] + \Pr[W=1|X=x,Y=y] = 1$. So $\theta$ is parameterized by $\Pr[W=1|X=x,Y=y]$, visualized in Figure \ref{fig:stable}. An example of communication:
\begin{description}
\item[1] Xayla writes she believes $W=1$ with probability $ > 0.5$;
\item[2] Yanis writes he believes $W=1$.
\end{description}

After round 1, Xayla's message implies $x=2$, since $\Pr[W=1|X=0]=\frac13$, $\Pr[W=1|X=1]=0.5$, and $\Pr[W=1|X=2]=\frac23$. Yanis infers $x=2$ from Xayla's message. His round 2 message then implies $y=1$, as only $\Pr[W=1|X=2,Y=1]=1$. Therefore, we can infer the stimulus was $\mathbf{x}=(x=2,y=1)$.

\end{example}

\paragraph{Histories are Hyperrectangles}\label{sec:rectangle} Given $\Pi$, any fixed history $h$, let $R_h$ denote all those inputs $\mathbf{x}$ that lead to this history. That is, knowing $h$ is equivalent to knowing $\mathbf{x}\in R_h$. An important observation is that when the policies are deterministic, $R_h$ is a hyperrectangle (\Cref{fig:stable}), stated as follows. Intuitively, this means in the communication protocol, agents' information is revealed \emph{separately}. 

\begin{observation}\cite{kushilevitz_nisan_1996}\label{lem:separating}
When the policies are deterministic, given any history $h$, there exists $B_i\subset \mathcal{S}_i,\forall i$ such that $R_h=\cap (X_i\in B_i)$ for all $\mathbf{x}$.
\end{observation}

We can write $\cap_i (X_i \in B_i)$ equivalently as $X_1 \in B_1, X_2 \in B_2, \dots$. For example, in \Cref{exa:setting}, after round 1, knowing $h^1$ equals knowing $X\in\{2\}, Y\in\{0,1,2\}$. After round 1, knowing $h^2$ equals knowing $X\in\{2\}, Y\in\{1\}$. Generally, we refer to a history $h$ as a hyperrectangle $\cap_i (X_i \in B_i)$ specified by subsets $B_i \subset \mathcal{S}_i$ for all $i$. 

As communication proceeds, more knowledge is gained. Let $\Pi^t$ denote the protocol truncated after $t$ rounds. $\Pi^t(\mathbf{X})=H^t$ is the random history variable, with realizations $h^t$. Given $\Pi$, each input $\mathbf{x}$ leads to a unique $h^t$. All possible $R_{h^t}$ partition the input space. As $t$ increases, the partition becomes more refined. Figures \ref{fig:partition} and \ref{fig:partition2} illustrate this progressive partitioning.

\begin{figure}
\definecolor{darkgreen}{rgb}{0.0, 0.5, 0.0}
\definecolor{lightgreen}{rgb}{0.0, 0.8, 0.0}
    \centering
    \begin{minipage}{0.2\textwidth}
        \centering
    \scalebox{0.6}{
    \begin{tikzpicture}
    \draw (0,0) rectangle (4,4);  

    \draw[color=darkgreen,line width=0.5mm] (0,0) rectangle (2,1.5);
    \draw[color=darkgreen,line width=0.5mm] (2,0) rectangle (4,1.5);
    \draw[color=darkgreen,line width=0.5mm] (0,1.5) rectangle (1.5,4);
    \draw[color=darkgreen,line width=0.5mm] (1.5,1.5) rectangle (4,4);
\end{tikzpicture} 
}
        \captionof{figure}{$\Pi^{t-1}(\mathbf{X})$} \label{fig:partition}
    \end{minipage}
    \begin{minipage}{0.2\textwidth}
        \centering
\scalebox{0.6}{
    \begin{tikzpicture}
    \draw[color=lightgreen,line width=0.3mm] (0,0) rectangle (4,4);  

    \draw[color=lightgreen,line width=0.3mm] (0,0) rectangle (0.8,1.5);

    \draw[color=lightgreen,line width=0.3mm] (2,0) rectangle (4,0.8);

    \draw[color=lightgreen,line width=0.3mm] (0,1.5) rectangle (1.5,2.75);
    \draw[color=lightgreen,line width=0.3mm] (0,2.75) rectangle (1.5,4);

    \draw[color=lightgreen,line width=0.3mm] (1.5,1.5) rectangle (2.75,4);
    \draw[color=lightgreen,line width=0.3mm] (2.75,1.5) rectangle (4,4);

    \draw[color=darkgreen,line width=0.5mm] (0,0) rectangle (2,1.5);
    \draw[color=darkgreen,line width=0.5mm] (2,0) rectangle (4,1.5);
    \draw[color=darkgreen,line width=0.5mm] (0,1.5) rectangle (1.5,4);
    \draw[color=darkgreen,line width=0.5mm] (1.5,1.5) rectangle (4,4);

\end{tikzpicture}
}
        \captionof{figure}{$\Pi^t(\mathbf{X})$} \label{fig:partition2}
    \end{minipage}
\end{figure}

\paragraph{Evolving Posterior Beliefs}\label{sec:belief}
Here are multiple posterior beliefs for $W$, conditioning on different knowledge. As communication proceeds, beliefs evolve. 
\begin{description}
\item [Global Belief] $\mathbf{q}_{h}$ denotes the posterior belief for target $W$ conditioning on history $h$, i.e., $\mathbf{q}_{h}(\omega)=\Pr_{\theta}[W=\omega|h]$.
\item [Local Belief]  $\mathbf{q}_{h,x_i}$ denotes the posterior belief for $W$ conditioning on $h$ and signal $x_i$, i.e., $\mathbf{q}_{h,x_i}(\omega)=\Pr_{\theta}[W=\omega|X_i=x_i,h]$.
\item [Ground Truth Belief] $\mathbf{q}_{\mathbf{x}}$ denotes the posterior belief for $W$ conditioning on all signals, i.e., $\mathbf{q}_{\mathbf{x}}(\omega)=\Pr_{\theta}[W=\omega|X_1=x_1,X_2=x_2,\cdots,X_n=x_n]$.
\end{description}

It is useful to consider an outsider observing the board but lacking private information. The outsider's belief is the global belief $\mathbf{q}_{h}(1) = \Pr[W=1|h]$.

Initially, $\mathbf{q}_{h^0}(1) = \Pr[W=1] = 0.5$. The local belief with $X=2$ is $\mathbf{q}_{h^0,X=2}(1) = \Pr[W=1|X=2] = \frac{2}{3}$. The local belief with $Y=1$ is $\mathbf{q}_{h^0,Y=1}(1) = \Pr[W=1|Y=1] = 0.5$. After round 1, Xayla's message implies $X=2$, so $\mathbf{q}_{h^1}(1)$ updates to $\Pr[W=1|X=2] = \frac{2}{3}$. The local belief with $Y=1$ updates to $\mathbf{q}_{h^1,Y=1}(1) = \Pr[W=1|X=2,Y=1] = 1$, equaling the ground truth belief. We will show it is possible for equal local and global beliefs that differ from the ground truth belief.

\paragraph{Rewards \& Costs}
$v(h,\omega) = \log \mathbf{q}_h(\omega)$ indicates the value of $h$ when $W=\omega$. It measures the accuracy of the posterior belief at $h$ by the log scoring rule \cite{winkler1969scoring,gneiting2007strictly}. We define $h$'s cost as its self-information $c(h)=\log\frac{1}{\Pr[h]}$ \cite{shannon1948mathematical}. We hypothesize that more surprising histories incur greater costs in perceiving and communicating. This is because it is efficient to assign longer coding for less likely items \cite{shannon1948mathematical}. 
\begin{description}
\item [Reward] At each round $t$, selected agent $i_t$'s reward will be the marginal value of $h^{t}$ conditioning on $h^{t-1}$, \[v(h^t,\omega)-v(h^{t-1},\omega)=\log \mathbf{q}_{h^t}(\omega)-\log \mathbf{q}_{h^{t-1}}(\omega)\] where $h^{t}=(h^{t-1},P_{i_t}(x_{i_t},h^{t-1}))$, if $W=\omega$ is revealed in the future. 
\item [Cost] Agent $i_t$ additionally costs
\[\beta(c(h^t)-c(h^{t-1}))=\beta \log\frac{1}{\Pr[h^t|h^{t-1}]}\footnote{$c(h^t)-c(h^{t-1}))=\log\frac{1}{\Pr[h^t|h^{t-1}]}$ because $\frac{\Pr[h^t]}{\Pr[h^{t-1}]}=\Pr[h^t|h^{t-1}]$ as $h^t$ refines $h^{t-1}$.}\]
\end{description}

Here are useful information-theoretic interpretations of the expected value, reward, and cost. Entropy $\en(X)$ measures the uncertainty of random variable $X$, conditional entropy measures leftover uncertainty, mutual information $\mi(X;Y)$ captures inter-variable correlations, and conditional mutual information assesses residual correlations beyond a third variable. A formal introduction of basic information-theoretic concepts is deferred to the appendix. 

\begin{itemize}
\item Expected Value = (Conditional) Mutual Information: $\E_{(h,\omega)\sim (H,W) }[v(h,\omega)]=\mi(H;W)$; given $h^{t-1}$, the expected marginal value, i.e., reward, of random $h^t\sim H^t$ is conditional mutual information $\mi(H^t;W|h^{t-1})$;
\item Expected Cost = (Conditional) Entropy: $\E_{h\sim H}[c(h)]=\en(H)$; given $h^{t-1}$, the expected marginal cost of random $h^t\sim H^t$ is conditional entropy $\en(H^t|h^{t-1})$. 
\end{itemize}

The distributed communication model shares key features with economic models of costly information acquisition \cite{stigler1961economics}. In these models, a decision-maker strategically acquires expensive information to maximize utility, similar to agents in our framework balancing accuracy rewards against communication costs. Later, we will show that the reward in the distributed communication model exactly matches the reward of selling information in a special prediction market.


\section{Stable Perception}

This section presents two definitions of stable perception based on the model. The first definition states that a stable state is reached when no agent can gain positive utility by choosing to continue, given that others choose to stop. However, validating this definition requires verifying across all possible agent policies, which can be complex. Therefore, we also provide a simpler, consensus-based definition. Under this definition, agents achieve stability if beliefs converge to consensus. While distinct, we draw connections between the equilibrium and consensus concepts. We then formally define multistable perceptions, where multiple distinct stable states exist, leading to diverse global beliefs. Finally, visual illustrations provide intuitive examples of stability and multistability.

\paragraph{Equilibrium-based Definition} We assume agents are myopic - concerned only with immediate utility gains. Under this assumption, a stable perception emerges when no single agent can gain by unilaterally deviating from the choice to stop.

\begin{definition} [$\beta$-equilibrium] Given an information structure $\theta$, a history $h$, $(\theta, h)$ achieves $\beta$-equilibrium if, for all agent $i$, her expected utility is non-positive for any policy, i.e., \[\mi(f(X_i,h);W|h)\leq \beta \en(f(X_i,h)|h)\] for all possible function $f$. 
\end{definition}

The $(\theta, h)$ defined above represents an equilibrium state because no single agent $i$ can gain positive utility by unilaterally deviating from the ``stop'' choice, given other agents remain stopped. Specifically, if agent $i$ continues while others are stopped, $i$ will be selected with probability $\frac1n$ at the next round. The expected immediate utility gain from adopting policy $P_i$ is then $\frac1n\left(\mi(P_i(X_i,h);W|h)-\beta \en(P_i(X_i,h)|h)\right)$ which is less than zero based on the above definition. 

\paragraph{Consensus-based Definition}

However, the equilibrium definition is difficult to verify in practice, as it requires checking across all possible agent policies. Therefore, we adopt an alternative consensus-based definition \cite{kong2023false} that is simpler to validate. Under this definition, a stable perception is reached when agents' beliefs converge to an approximate consensus in expectation. We establish connections between this consensus definition and the prior equilibrium concept.

\begin{definition} [$\epsilon$-consensus] \cite{kong2023false} Given an information structure $\theta$, a history $h$, $(\theta, h)$ achieves $\epsilon$-consensus if \[\mi(X_i;W|h)\leq \epsilon,\forall i\] 
\end{definition}


We briefly explain why the above definition represents consensus among agents. Mutual information equals the expected KL divergence between posteriors and priors $\mi(X_i;W|h)=\E_{x_i\sim X_i|h} D_{KL}(\mathbf{q}_{h,x_i},\mathbf{q}_h)=\sum_{x_i} \Pr[X_i=x_i|h] D_{KL}(\mathbf{q}_{h,x_i},\mathbf{q}_h)$. KL divergence\footnote{KL divergence measures the difference between two probability distributions $\mathbf{p}$ and $\mathbf{q}$. $
D_{KL}(\mathbf{p},\mathbf{q}) = \sum_{\omega}\mathbf{p}(\omega)\log\frac{\mathbf{p}(\omega)}{\mathbf{q}(\omega)}$.} quantifies how much additional signal $x_i$ shifts belief for $W$, given $h$. In essence, it captures the amount of information shared by $X_i$ and $W$, conditioning on $h$.

So the above definition implies local beliefs approximately equal the global belief in expectation - all reach approximate consensus.

\begin{propositionrep}[Connecting $\beta$-equilibrium \& $\epsilon$-consensus]
Given any information structure $\theta$, any history $h$, if $(\theta, h)$ is $\beta$-equilibrium and $\beta\leq b(\epsilon)$\footnote{$b(\epsilon)=\frac{1}{16}\epsilon^2\frac{1}{\log\frac{1}{E^{-1}(\epsilon)}}$ where $E^{-1}(\epsilon)\leq 0.5$ and is the solution of $x\log x+(1-x)\log (1-x)=- \epsilon$.}, then $(\theta, h)$ is $\epsilon$-consensus. 
\end{propositionrep}
We prove this by contradiction. The key observation is that if $(\theta, h)$ is not $\epsilon$-consensus, one agent has a policy with a large reward $\geq 2b(\epsilon)$ and a small cost $\leq 2$. Thus, with $\beta\leq b(\epsilon)$, $\beta$-equilibrium is a special $\epsilon$-consensus. It suffices to analyze $\epsilon$-consensus's properties.

\begin{appendixproof}

We prove the results by contradiction. If $(\theta, h)$ is not $\epsilon$-consensus, based on the compression idea in work \cite{aaronson2005complexity}, the prior study proves the following results.  

\begin{lemma}\cite{kong2023false}
When $I(X_i;W|h)>\epsilon$, there exists $P_i$ which maps everything to $\{\text{high, medium or low }\}$ such that \[I(P_i(X_i,h);W|h)> 2b(\epsilon) \] where $E^{-1}(\epsilon)\leq 0.5$ and is the solution of $x\log x+(1-x)\log (1-x)=- \epsilon$.
\end{lemma}

As $P_i$'s range is $\{\text{high, medium or low }\}$, agent $i$'s expected information cost is less than $\beta \cdot 2$ bits. When agent $i$ adopts $P_i$, her expected payment is 
\begin{align*}
I(P_i(X_i,h);W|h)>2b(\epsilon)\geq 2\beta
\end{align*}

Thus, $(\theta, h)$ is not a $\beta$-equilibrium. We finished the proof. 
\end{appendixproof}

\paragraph{Multistable Perceptions} Once we have established the definition of stable perception, we are ready to give the multistable perception. Intuitively, we perceive more than one stable perception of a stimulus if the stable perceptions lead to divergent beliefs for $W$. We use an additional parameter $d$ to quantify the divergence between the stable perceptions.  

\begin{definition}[Multistable Perception]
An stimulus $\mathbf{x}$ that is drawn from $\theta$ has $(\epsilon,d)$-multistable perception if there exists two history of public messages $h,h'\ni \mathbf{x}$ such that both $(\theta, h)$ and $(\theta,h')$ achieve $\epsilon$-consensus, and \[D_{TV}(\mathbf{q}_h,\mathbf{q}_{h'})\geq d.\]
\end{definition}

In the above definition, we use total variation distance $D_{TV}(\mathbf{p},\mathbf{q})=\frac12\sum_{\omega}|\mathbf{p}(\omega)-\mathbf{q}(\omega)|$ instead of KL divergence for a technical reason: the total variation distance satisfies the triangle inequality while KL does not. 

\paragraph{Visual Illustrations} \Cref{exa:setting} presents a visual interpretation of the above concepts.

\begin{figure}[ht]
   \centering
   \begin{subfigure}[t]{0.4\textwidth} 
      \centering
      \includegraphics[height=.5\textwidth]{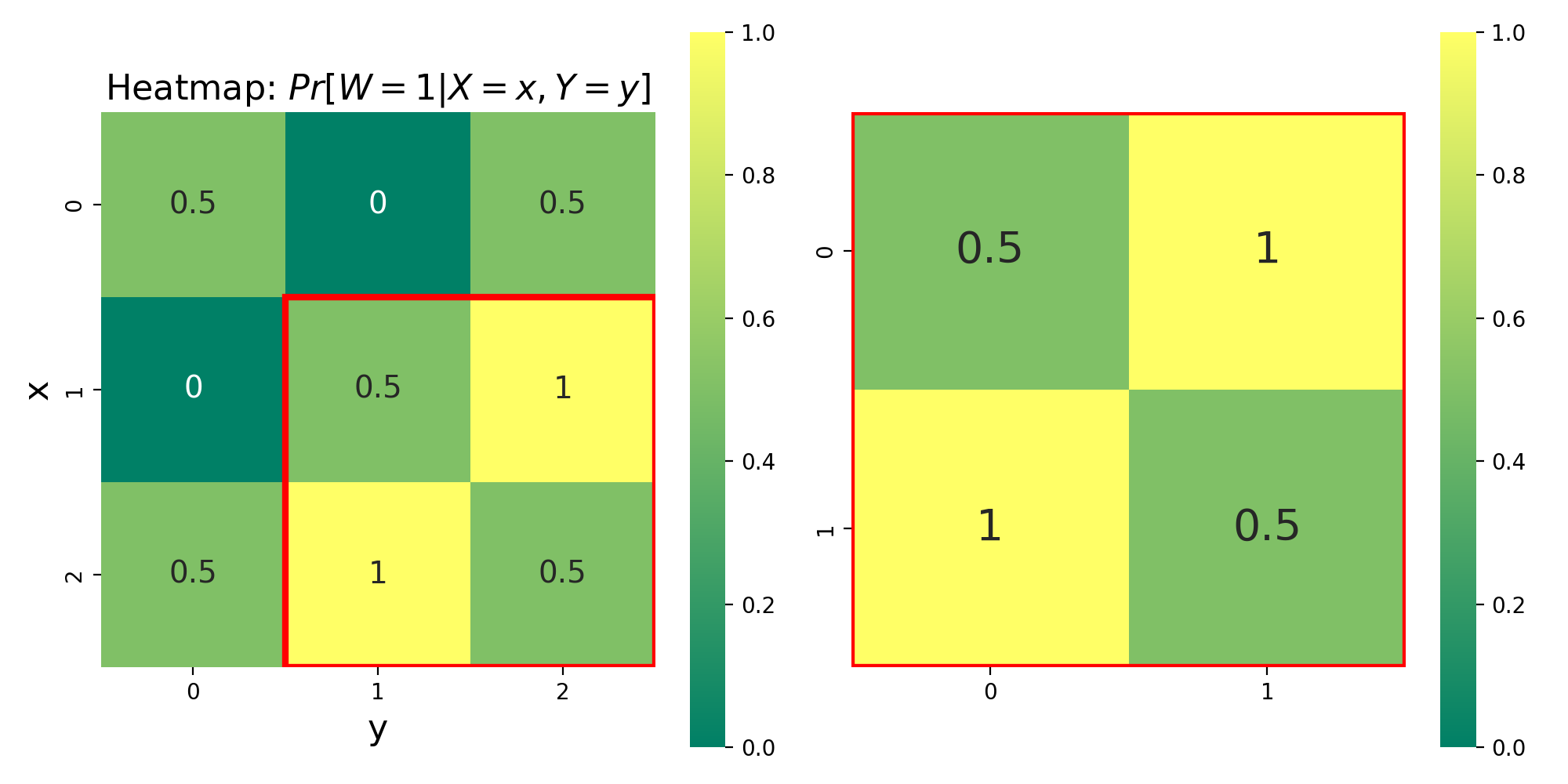}
      \caption{$R_h:X\in\{1,2\},Y\in\{1,2\}$ }
      \label{fig:heatmap1}
   \end{subfigure}
   \hfill 
   \begin{subfigure}[t]{0.4\textwidth}
      \centering
      \includegraphics[height=.5\textwidth]{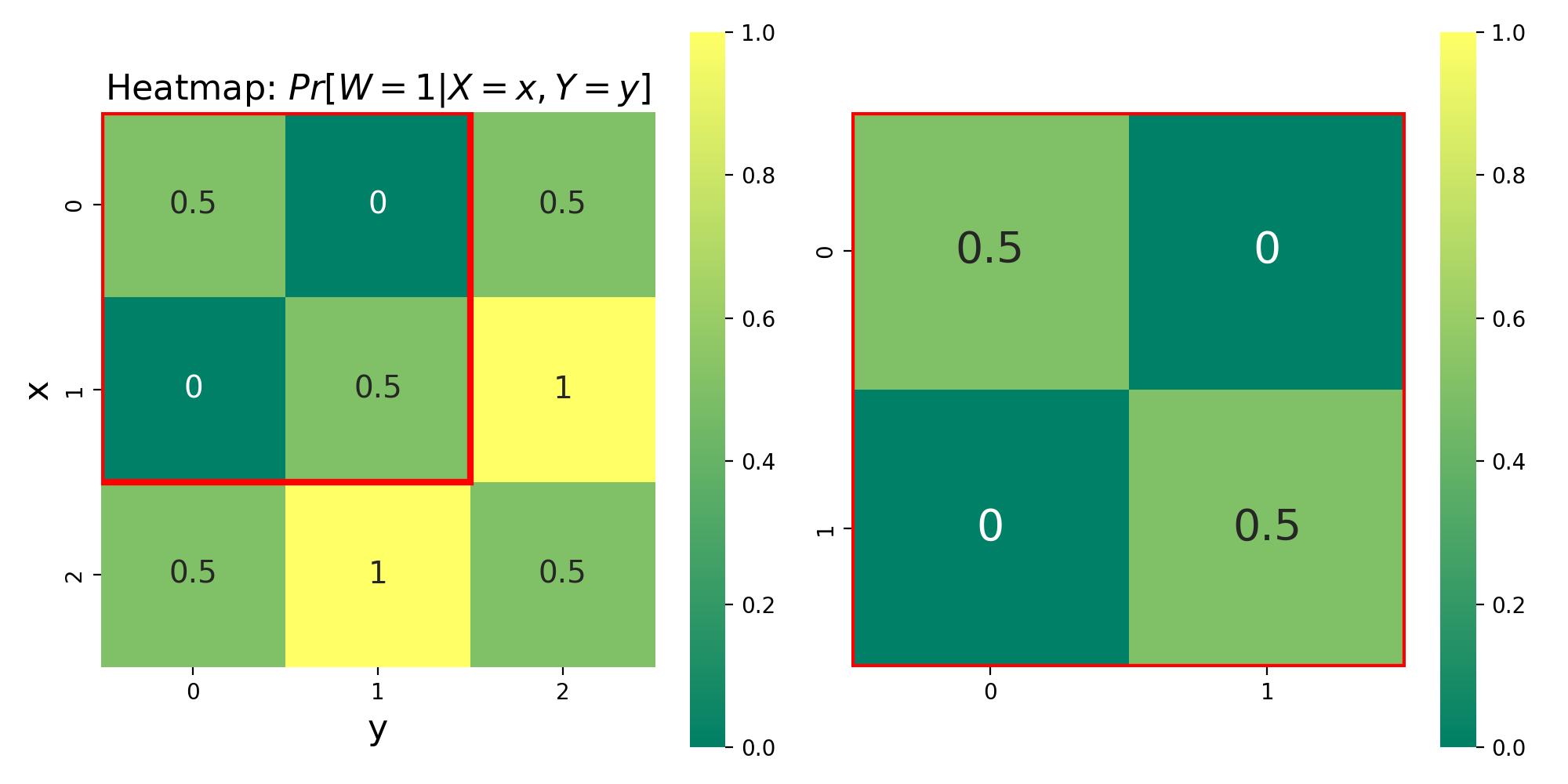}
      \caption{$R_h:X\in\{0,1\},Y\in\{0,1\}$}
      \label{fig:heatmap2}
   \end{subfigure}
   \hfill 
   \begin{subfigure}[t]{0.4\textwidth}
      \centering
      \includegraphics[height=.5\textwidth]{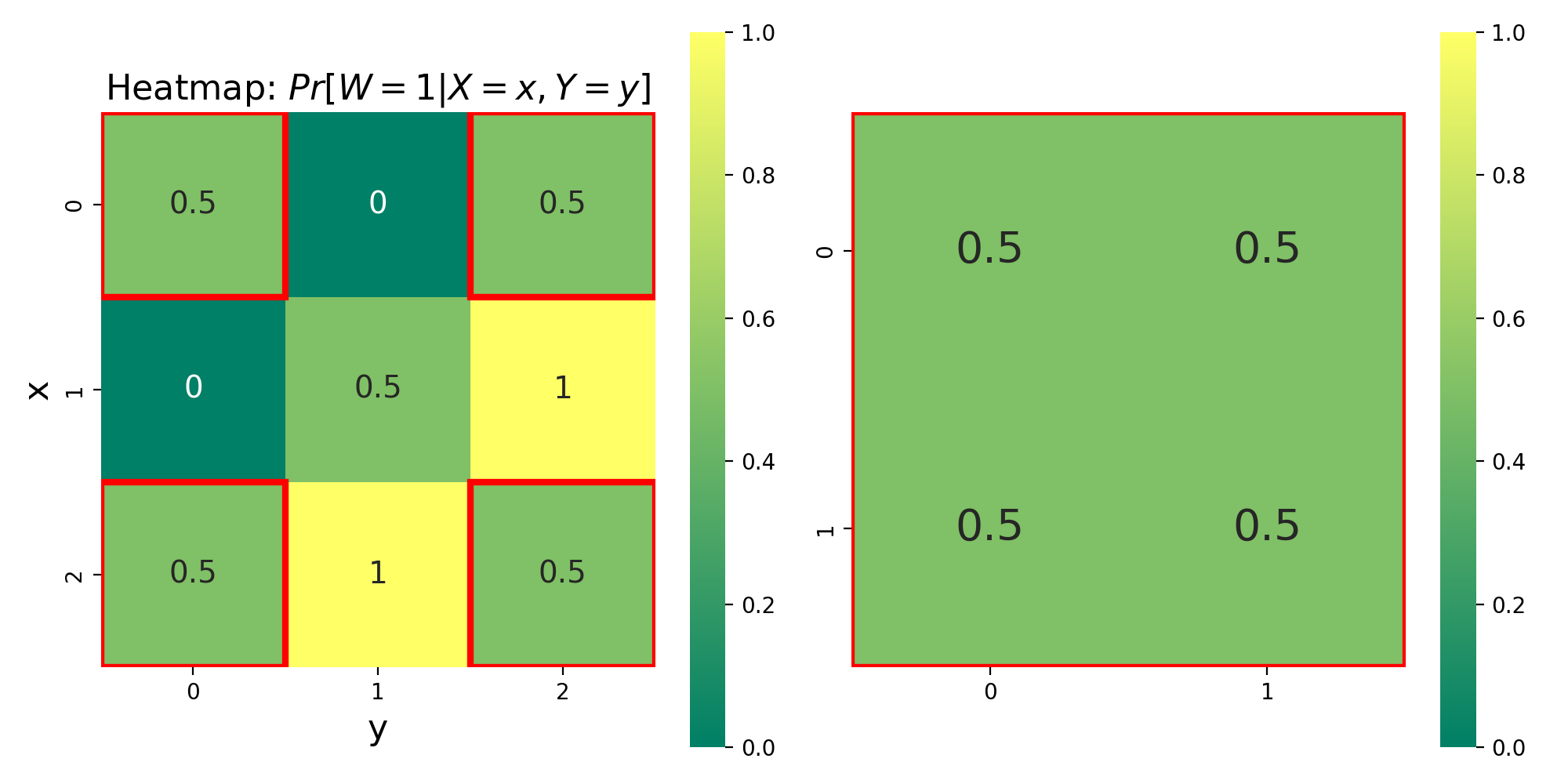}
      \caption{$R_h:X\in\{0,2\},Y\in\{0,2\}$}
      \label{fig:heatmap3}
   \end{subfigure}
   \caption{\textbf{Examples of Stable Perceptions}}
   \label{fig:stable}
\end{figure}

\begin{example}[Visual Illustrations]
Because $W$ in \Cref{exa:setting} is a binary event, we shorthand posterior $\mathbf{q}(1)$ as $q$.
\begin{claim}
In \Cref{exa:setting}, 
\begin{itemize}
\item History $h$ is rectangle $R_h=X\in B_x, Y\in B_y$
\item Ground truth belief $q_{x,y}=\Pr[W=1|X=x,Y=y]$ equals $(x,y)$'s heatmap value
\item Global belief $q_h$ is the average of values in $R_h$
\item Local belief $q_{h,x}$ is the average of $x^\text{th}$ row in $R_h$
\item Local belief $q_{h,y}$ is the average of $y^\text{th}$ column in $R_h$
\end{itemize}

\end{claim}

With the visualization,
$\mi(X; W|h)=\E_{x\sim X|h} D_{KL}(\mathbf{q}_{h,x},\mathbf{q}_h)$
interprets as the variance - expected divergence between individual and average - of the row averages. Analogously, $\mi(Y; W|h)$ is the variance of the column averages.

Figure \ref{fig:stable} shows three examples of perfect consensuses (0-consensus), with zero variance for row/column averages. Values in perfect consensus $h$ can have high (\ref{fig:heatmap1}, \ref{fig:heatmap2}) or low (\ref{fig:heatmap3}) variance. Later we show variance relates to ``regret'' - the amount of unaggregated information.

With $x=y=1$, at least two perfect consensuses exist, leading to distinct perceptions $q_h=0.75$ (\ref{fig:heatmap1}) or $q_h=0.25$ (\ref{fig:heatmap2}). This connects to the phenomenon of multistable perception.  
\end{example}

\section{Impact of Relationship Among Components}

This section\footnote{Due to space limitations, most proofs in this section and the later sections are deferred to the appendix.} demonstrates how component relationships influence stable perception formation, showing substitutes lead to monostability. A core concept is consensus regret - the amount of unaggregated information. True consensus has zero regrets. 

\begin{definition}[Regret of Stable Perceptions]
Regret of history $h$ is the conditional mutual information between all information and $W$, conditioning on $h$, i.e., $\mi(\mathbf{X};W|h)$.
\end{definition}

In Example \ref{exa:setting}, regret of history $h$ is $\mi(X,Y; W|h)=\E_{x, y\sim X,Y|h} D_{KL}(\mathbf{q}_{x,y},\mathbf{q}_h)$
that interprets as the variance - expected divergence between individual and average - of the values in $R_h$. High regret corresponds to diverging patterns (\ref{fig:heatmap1}, \ref{fig:heatmap2}). Low regret corresponds to converging patterns (\ref{fig:heatmap3}).

\subsection{Substitutes $\Rightarrow$ Monostable Perception}

This section illustrates that substitutes lead to low-regret stable perceptions, thus inducing monostable perception. 

\paragraph{Substitutes $\Rightarrow$ Low-Regret Stable Perceptions} We introduce a special information structure, information substitutes: $X_1,X_2,\cdots, X_n$ are independent conditioning on $W$. Prior work proved that under information substitutes, all approximate consensuses are approximately true, i.e., have low regrets \cite{kong2023false,frongillo2023agreement}\footnote{An alternative model is also studied to show that information aggregation in social learning can be inefficient with complementary information \cite{liang2020complementary}.}. The key idea is that the information substitutes satisfy the subadditivity $\mi(\mathbf{X};W|h)\leq \sum_i \mi(X_i;W|h)$ for all $h$. This property upper-bounds the regret by the sum of marginal regret of all agents.

\paragraph{Low-Regret Stable Perceptions $\Rightarrow$ Monostable Perception} 
The following theorem states under substitutes, two $\epsilon$-consensuses with non-negligible overlapping will lead to approximately the same global beliefs, resulting in a monostable perception. 

\begin{theorem}[Substitutes $\Rightarrow$ Monostable Perception]\label{thm:sub}
When $X_1,X_2,\cdots, X_n$ are independent conditioning on $W$, given $h,h'$, if both $(\theta, h)$ and $(\theta, h')$ reach perfect consensus, then $\mathbf{q}_h=\mathbf{q}_{h'}=\mathbf{q}_{\mathbf{x}}$; if both $(\theta, h)$ and $(\theta, h')$ achieve $\epsilon$-consensus and both $\Pr[h|h']$ and $\Pr[h'|h]$ are strictly greater than $\frac{2 n\epsilon}{d^2}$, then $D_{TV}(\mathbf{q}_h,\mathbf{q}_{h'})< d$.
\end{theorem} 

Under substitutes, in perfect consensuses, information is fully aggregated, aligning consensus with ground truth belief. This prohibits divergent interpretations, forcing monostable perception. In $\epsilon$-consensuses, information is almost fully aggregated. Non-negligibly overlapping $\epsilon$-consensuses cannot be highly diverse. 


\begin{appendixproof}[Proof of \Cref{thm:sub}]

We prove a slightly stronger lemma, whose results will also be used to analyze the switching cost later. 

\begin{lemmarep}\label{lem:general}
Given any pair of history public messages $h,h'$, if both $(\theta, h)$ and $(\theta, h')$ achieve $\epsilon$-consensus and both $\Pr[h|h']$ and $\Pr[h'|h]$ are strictly greater than $\min\{\frac{2 \max\{ \mi(\mathbf{X};W|h), \mi(\mathbf{X};W|h')\}}{d^2},1-\frac{d}{2}\}$, then $D_{TV}(\mathbf{q}_h,\mathbf{q}_{h'})< d$.
\end{lemmarep} 

\begin{proof}[Proof of \Cref{lem:general}]
We prove the result by contradiction. 

When $D_{TV}(\mathbf{q}_h,\mathbf{q}_{h'})\geq d$, then either $D_{TV}(\mathbf{q}_{h'\cap h},\mathbf{q}_h)>\frac{d}{2}$ or $D_{TV}(\mathbf{q}_{h'\cap h},\mathbf{q}_{h'})>\frac{d}{2}$. 

Without loss of generality, we assume $D_{TV}(\mathbf{q}_{h'\cap h},\mathbf{q}_h)>\frac{d}{2}$. 
\begin{align*}
D_{TV}(\mathbf{q}_{h'\cap h},\mathbf{q}_h) = & \frac12 ||\mathbf{q}_{h'\cap h}-\mathbf{q}_h||_1\\
=&\frac12 ||\mathbf{q}_{h'\cap h}-(\Pr[h'|h]\mathbf{q}_{h'\cap h}+(1-\Pr[h'|h])\mathbf{q}_{\neg h'\cap h})||_1\\
=& \frac12 (1-\Pr[h'|h])||\mathbf{q}_{h'\cap h}- \mathbf{q}_{\neg h'\cap h})||_1\\
\leq & 1-\Pr[h'|h]
\end{align*}

This leads to a trivial bound $\Pr[h'|h]\leq 1-\frac{d}{2}$. We then prove the non-trivial bound. 

By Fubini's theorem, \begin{align*}
 \mi(\mathbf{X};W|h)=\E_{\mathbf{x}\sim\mathbf{X}|h} D_{KL}(\mathbf{q}_{\mathbf{x}},\mathbf{q}_h)=\Pr[h'|h] \E_{\mathbf{x}\sim\mathbf{X}|h'\cap h} D_{KL}(\mathbf{q}_{\mathbf{x}},\mathbf{q}_h) + (1-\Pr[h'|h]) \E_{\mathbf{x}\sim\mathbf{X}|\neg h'\cap h} D_{KL}(\mathbf{q}_{\mathbf{x}},\mathbf{q}_h)
\end{align*}

Therefore, $\Pr[h'|h] \E_{\mathbf{x}\sim\mathbf{X}|h'\cap h} D_{KL}(\mathbf{q}_{\mathbf{x}},\mathbf{q}_h)\leq  \mi(\mathbf{X};W|h)$. To upper bound $\Pr[h'|h]$, it's left to lower bound $ \E_{\mathbf{x}\sim\mathbf{X}|h'\cap h} D_{KL}(\mathbf{q}_{\mathbf{x}},\mathbf{q}_h)$. 

We first lower bound the KL divergence between $\mathbf{q}_{h'\cap h}$ and $\mathbf{q}_h$. Because for any $\mathbf{p},\mathbf{q}$, $D_{TV}(\mathbf{p},\mathbf{q})\leq \sqrt{\frac12 D_{KL}(\mathbf{p},\mathbf{q})}$, we then have $D_{KL}(\mathbf{q}_{h'\cap h},\mathbf{q}_h)>2 (D_{TV}(\mathbf{q}_{h'\cap h},\mathbf{q}_h))^2\geq 2 (\frac{d}{2})^2$. 

We then lower bound the expected KL divergence between $\mathbf{q}_{\mathbf{x}}, \mathbf{x}\in h\cap h'$ and $\mathbf{q}_h$, $ \E_{\mathbf{x}\sim\mathbf{X}|h'\cap h} D_{KL}(\mathbf{q}_{\mathbf{x}},\mathbf{q}_h)$, via the convexity of KL divergence. 

\begin{align*}
& \E_{\mathbf{x}\sim\mathbf{X}|h'\cap h} D_{KL}(\mathbf{q}_{\mathbf{x}},\mathbf{q}_h)\\
\geq & D_{KL}(\mathbf{q}_{h'\cap h},\mathbf{q}_h) \geq \frac{d^2}{2}
\end{align*}

Combined the above results, we have $\Pr[h'|h] \leq \frac{2  \mi(\mathbf{X};W|h)}{d^2}$.
\end{proof}

Due to subadditivity, we have the expected KL divergence between $\mathbf{q}_{\mathbf{x}}, \mathbf{x}\in h$ and $\mathbf{q}_h$ is bounded. Formally, $ \E_{\mathbf{x}\sim\mathbf{X}|h} D_{KL}(\mathbf{q}_{\mathbf{x}},\mathbf{q}_h)=\mi(\mathbf{X};W|h)\leq n\epsilon$. The results follow directly from \Cref{lem:general}.

\end{appendixproof}

\subsection{Low Complexity $\Rightarrow$ Monostable Perception}

In general, we provide a complexity measure that measures the worst-case regret of the stable perceptions and present an analogous result to \Cref{thm:sub}. We also show that the complexity measure satisfies a direct-sum property that the complexity of m copies is m times the complexity of a single copy.

\begin{definition}[Gestalt Complexity]
An $\epsilon$-consensus $(\theta,h)$'s Gestalt complexity is defined as regret $g(\theta, h,\epsilon)=\mi(\mathbf{X};W|h)$. An stimulus $\mathbf{x}$'s Gestalt complexity is defined as
$g(\theta,\mathbf{x},\epsilon)=\max_{R_h\ni \mathbf{x},h\text{ is $\epsilon$-c}} \mi(\mathbf{X};W|h)$. $\theta$'s Gestalt complexity is defined as
$g(\theta,\epsilon)=\max_{h\text{ is $\epsilon$-c}} \mi(\mathbf{X};W|h)$.
\end{definition}

$\epsilon$-c is a shorthand of $\epsilon$-consensus. A stimulus has low complexity if all $\epsilon$-consensuses that contain the stimulus have a low regret. An information structure has low complexity if its all $\epsilon$-consensuses have a low regret. Thus, under low complexity, all $\epsilon$-consensuses are approximately true as well. We could adopt the proof of \Cref{thm:sub} to directly prove the following results.

\begin{theorem}[Low Complexity $\rightarrow$ Monostable Perception]\label{thm:lowcomplexity}
Given any stimulus $\mathbf{x}$ that is drawn from $\theta$, given any pair of history public messages $R_h,R_{h'}\ni\mathbf{x}$, if both $(\theta, h)$ and $(\theta, h')$ are $\epsilon$-consensuses and both $\Pr[h|h']$ and $\Pr[h'|h]$ are strictly greater than $\frac{2 g(\theta,\mathbf{x},\epsilon)}{d^2}$, then $D_{TV}(\mathbf{q}_h,\mathbf{q}_{h'})< d$.
\end{theorem}


\paragraph{Direct-sum Property}  The direct-sum property of a complexity measure concerns whether the complexity of m copies as m times the
complexity of a single copy. $\theta^A\oplus\theta^B$ is the direct sum of two information structures where \[\Pr_{\theta^A\oplus\theta^B}[\mathbf{X}^A,\mathbf{X}^B,W^A,W^B]=\Pr_{\theta^A}[\mathbf{X}^A,W^A] \Pr_{\theta^B}[\mathbf{X}^B,W^B].\] $h^A$ is the partial knowledge about $\mathbf{X}_A$. $h^B$ is the partial knowledge about $\mathbf{X}_B$. We can see $h^A,h^B$ as subsets of $\mathbf{X}_A,\mathbf{X}_B$'s realization spaces respectively. $h^A\oplus h^B$ denotes the joint knowledge $h^A\cap h^B$. 

\begin{propositionrep}[Direct-sum]\label{thm:directsum}
Given information structure $\theta^A\oplus\theta^B$, a stimulus $(\mathbf{x}^A,\mathbf{x}^B)$ and history $h^A\oplus h^B$, we have 
\[g(\theta^A\oplus\theta^B,h^A\oplus h^B,\epsilon)=g(\theta^A,h^A,\epsilon)+g(\theta^B,h^B,\epsilon)\]
\[g(\theta^A\oplus\theta^B,(\mathbf{x}^A,\mathbf{x}^B),\epsilon)=g(\theta^A,\mathbf{x}^A,\epsilon)+g(\theta^B,\mathbf{x}^B,\epsilon)\]
\end{propositionrep}

We mainly use the fact that $\log ab=\log a+\log b$ to separate the mutual information measure. We defer the full proof to the appendix. 

\begin{appendixproof}
First, because $\Pr_{\theta^A\oplus\theta^B}[X^A_i,W^A,W^B|h^A\oplus h^B]=\Pr_{\theta^A}[X^A_i,W^A|h^A]\Pr_{\theta^B}[W^B|h^B]$, 
\begin{align*}
& \mi(X^A_i;W^A,W^B|h^A\oplus h^B)\\
= & \sum_{x^A_i,\omega^A,\omega^B} \Pr_{\theta^A\oplus\theta^B}[X^A_i,W^A,W^B|h^A\oplus h^B] \log \frac{\Pr_{\theta^A\oplus\theta^B}[X^A_i,W^A,W^B|h^A\oplus h^B]}{\Pr_{\theta^A\oplus\theta^B}[X^A_i|h^A\oplus h^B]\Pr_{\theta^A\oplus\theta^B}[W^A,W^B|h^A\oplus h^B]}\\
= & \sum_{x^A_i,\omega^A,\omega^B} \Pr_{\theta^A\oplus\theta^B}[X^A_i,W^A,W^B|h^A\oplus h^B] \log \frac{\Pr_{\theta^A}[X^A_i,W^A|h^A]\Pr_{\theta^B}[W^B|h^B]}{\Pr_{\theta^A}[X^A_i|h^A]\Pr_{\theta^A}[W^A|h^A]\Pr_{\theta^B}[W^B|h^B]}\\
= & \sum_{x^A_i,\omega^A,\omega^B} \Pr_{\theta^A\oplus\theta^B}[X^A_i,W^A,W^B|h^A\oplus h^B] \log \frac{\Pr_{\theta^A}[X^A_i,W^A|h^A]}{\Pr_{\theta^A}[X^A_i|h^A]\Pr_{\theta^A}[W^A|h^A]}\\
= & \mi(X^A_i;W^A|h^A)
\end{align*}

Analogously, \begin{align*}
& \mi(X^B_j;W^A,W^B|h^A\oplus h^B)\\
= & \mi(X^B_j;W^B|h^B)
\end{align*}

Therefore, $h^A\oplus h^B$ is $\epsilon$-consensus if and only if both $h^A$ and $h^B$ are $\epsilon$-consensus. 

Additionally, we have \[\Pr_{\theta^A\oplus\theta^B}[\mathbf{X}^A,\mathbf{X}^B,W^A,W^B|h^A\oplus h^B]=\Pr_{\theta^A}[\mathbf{X}^A,W^A|h^A] \Pr_{\theta^B}[\mathbf{X}^B,W^B| h^B].\]

\begin{align*}
& \mi(\mathbf{X}^A,\mathbf{X}^B;W^A,W^B|h^A\oplus h^B)\\
= & \sum_{\mathbf{x}^A\in h^A,\mathbf{x}^B\in h^B,\omega^A,\omega^B} \Pr_{\theta^A\oplus\theta^B}[\mathbf{X}^A,\mathbf{X}^B,W^A,W^B|h^A\oplus h^B]\\
&\log(\frac{\Pr_{\theta^A\oplus\theta^B}[\mathbf{X}^A,\mathbf{X}^B,W^A,W^B|h^A\oplus h^B]}{\Pr_{\theta^A\oplus\theta^B}[\mathbf{X}^A,\mathbf{X}^B|h^A\oplus h^B]\Pr_{\theta^A\oplus\theta^B}[W^A,W^B|h^A\oplus h^B]})\\
= & \sum_{\mathbf{x}^A\in h^A,\mathbf{x}^B\in h^B,\omega^A,\omega^B} \Pr_{\theta^A\oplus\theta^B}[\mathbf{X}^A,\mathbf{X}^B,W^A,W^B|h^A\oplus h^B]\\
&\log(\frac{\Pr_{\theta^A}[\mathbf{X}^A,W^A|h^A]\Pr_{\theta^B}[\mathbf{X}^B,W^B|h^B]}{\Pr_{\theta^A}[\mathbf{X}^A|h^A]\Pr_{\theta^A}[W^A]\Pr_{\theta^B}[\mathbf{X}^B|h^B]\Pr_{\theta^B}[W^B|h^B]})\\
= & \sum_{\mathbf{x}^A\in h^A,\mathbf{x}^B\in h^B,\omega^A,\omega^B} \Pr_{\theta^A\oplus\theta^B}[\mathbf{X}^A,\mathbf{X}^B,W^A,W^B|h^A\oplus h^B]\\
&\left(\log(\frac{\Pr_{\theta^A}[\mathbf{X}^A,W^A|h^A]}{\Pr_{\theta^A}[\mathbf{X}^A|h^A]\Pr_{\theta^A}[W^A|h^A]})+\log(\frac{Pr_{\theta^B}[\mathbf{X}^B,W^B|h^B]}{\Pr_{\theta^B}[\mathbf{X}^B|h^B]\Pr_{\theta^B}[W^B|h^B]})\right)\\
=&\mi(\mathbf{X}^A;W^A|h^A) + \mi(\mathbf{X}^B;W^B|h^B)\\
\end{align*}

Therefore, $g(\theta^A\oplus\theta^B,h^A\oplus h^B,\epsilon)=g(\theta^A,h^A,\epsilon)+g(\theta^B,h^B,\epsilon)$. 


Because $g(\theta,\mathbf{x},\epsilon)=\max_{R_h\ni\mathbf{x},h\text{ is $\epsilon$-consensus}} g(\theta, h,\epsilon) $, $g(\theta,\epsilon)=\max_{\mathbf{x}} g(\theta,\mathbf{x},\epsilon)$, and $g(\theta,\mathbf{x})=\max_{R_h\ni\mathbf{x}} g(\theta, h) $, $g(\theta)=\max_{\mathbf{x}} g(\theta,\mathbf{x})$, other direct-sum results also follow.

\end{appendixproof}

\subsection{Order Effect}

One intriguing phenomenon in multistable perceptions is that the perception depends on the order view perceives the image. The boundary and chaos-order examples in \Cref{fig:multistable} are two examples. The communication model illustrates a stimulus with multiple high regret stable perceptions can lead to diverse perceptions under different orders.

\begin{observation}
There exists communication protocol $\Pi$ and stimulus $\mathbf{x}$ where different $\mathbf{i}$ leads to stable perceptions with total variation distance $\geq 0.5$.
\end{observation}

Recall that $\mathbf{i}$ is the communication order of agents. The total variation distance is between 0 to 1. Thus, $\geq 0.5$ implies quite diverging perceptions. 

\begin{example}[Order Effect]
We use the information structure $\theta$ in \Cref{exa:setting}. The stimulus is $x=1,y=1$. We then describe Xayla and Yanis's policies. In both protocols, at all rounds, Xayla will write whether her current belief for $W=1$ is $\geq 0.6$ or $< 0.6$. Yanis will write whether his current belief for $W=1$ is $\geq 0.5$ or $< 0.5$. 

\paragraph{Communication A} Yanis writes in the first round, and Xayla writes in the second round. 
\begin{description}
\item[1] Yanis writes that he believes $W$ happens with prob $\geq 0.5$; 
\item[2] Xayla writes that she believes $W$ happens with prob $\geq 0.6$.
\end{description}

Yanis's round 1 message implies that $y$ is 1 or 2 because $\Pr[W=1|Y=0]=\frac13$, $\Pr[W=1|Y=1]=0.5$ and $\Pr[W=1|Y=2]=\frac23$. Xayla infers this and her round 2 message implies that $x$ is 1 or 2 because $\Pr[W=1|X=1,Y\in\{1,2\}]=\Pr[W=1|X=2,Y\in\{1,2\}]=0.75$, while $\Pr[W=1|X=0,Y\in\{1,2\}]=0.25$. After two rounds, they reach a perfect consensus $h: X\in\{1,2\},Y\in\{1,2\}$ where $q_h = 0.75$ (\Cref{fig:heatmap1}). 

\paragraph{Communication B} Xayla writes in the first round, and Yanis writes in the second round. 
\begin{description}
\item[1] Xayla writes that she believes $W$ happens with prob $<0.6$;
\item[2] Yanis writes that he believes $W$ happens with prob $<0.5$.
\end{description}

Xayla's round 1 message implies that $x$ is 0 or 1, because $\Pr[W=1|X=0]=\frac13$, $\Pr[W=1|X=1]=0.5$ and $\Pr[W=1|X=2]=\frac23$. Yanis infers this and his round 2 message implies that $y$ is 0 or 1 because $\Pr[W=1|X\in\{0,1\},Y=0]=\Pr[W=1|X\in\{0,1\},Y=0]=0.25$, while $\Pr[W=1|X\in\{0,1\},Y=2]=0.75$. After two rounds, they reach a perfect consensus $h: X\in\{0,1\},Y\in\{0,1\}$ where $q_h = 0.25$ (\Cref{fig:heatmap2}). 
\end{example}

Therefore, when multiple high regret stable perceptions exist, order effects may arise.

\section{Switch or Not?}

In this section, we will formally provide an information-theoretic approach to quantify the difficulty of switching, which provides a distinguishment between the easy-switching multistable perceptions and hard-switching multistable perceptions. 

\paragraph{Switching = Loop(Perceiving+Forgetting)}

Two possible explanations for switching are:

\begin{itemize}
\item Perception Disagreements: Agents reach $h^t$ in round $t$ with perception $\mathbf{q}_{h^t}$ and $h^{t+1}$ in round $t+1$ with perception $\mathbf{q}_{h^{t+1}}$ where $D_{TV}(\mathbf{q}_{h^{t}},\mathbf{q}_{h^{t+1}})>d$. 
\item Loop(Perceiving+Forgetting): After agents reach a stable $h$, if they want to switch, they will choose to forget and return to $\hat{h}\supset h$, and then go to another stable $h'\subset \hat{h}$. The process can be repeated. 
\end{itemize}

Prior studies showed that disagreements will not happen a lot with high probability, given that these perceptions are Bayesian posteriors \cite{aaronson2005complexity,kong2023false}. Intuitively, frequent disputes would allow an outsider to gain non-negligible information. As total information is bounded only by the entropy of $W$, substantial disagreements cannot persist with high probability. We adapt this analysis to our multistability setting, showing stable states with frequent divergence could not emerge under reasonable assumptions. 

\begin{propositionrep}
For any $\theta$ and $\Pi$, the number of disagreements is bounded by $\frac{\en(W)}{2d^2\delta}$ with probability $1-\delta$.
\end{propositionrep}

Frequent switching contradicts the improbability of disagreements between stable beliefs. Yet in practice, we can switch freely given enough time. This motivates modeling switching as a loop of perceiving and forgetting, rather than continuous disagreement.

\begin{appendixproof}
We use $A_L$ to denote the set of all possible instances $\mathbf{x}$ which has $>L$ disagreements with communication protocol $\Pi$. 

We pick sufficient large $t_L$ such that all instances in $A_L$ have already experienced $L$ disagreements after $t_L$ rounds. If an instance has experienced $<t_L$ communication rounds, we define the perceptions after its stopping round as the perception at the stopping round. 

\begin{align*}
\mi(H^{t_L};W) =& \mi(H^1;W)+\mi(H^2;W|H^1)+\cdots+\mi(H^{t_L};W|H^1,H^2,\cdots,H^{t_L-1})\\
= & \E[D_{KL}(\mathbf{q}_{h^1},\mathbf{q})]+\E[D_{KL}(\mathbf{q}_{h^2},\mathbf{q}_{h^1})] +\cdots \E[D_{KL}(\mathbf{q}_{h^{t_L}},\mathbf{q}_{h^{t_L-1}})]\\
\geq & \Pr[\mathbf{X}\in A_L] 2d^2 L
\end{align*}

Because the mutual information is less than the entropy $\mi(H^{t_L};W)\leq \en(W)$, we have $L\leq \frac{\en(W)}{\Pr[\mathbf{X}\in A_L] 2d^2}$. 

$\Pr[\mathbf{X}\in A_L]$ is the probability that $\mathbf{x}$ has experienced $>L$ disagreements. Thus, the number of disagreements is bounded by $\frac{\en(W)}{2d^2\delta}$ with probability $1-\delta$.  

\end{appendixproof}

\paragraph{Switching Cost}

We define the switching cost as the sum of forgetting cost and perceiving cost. 

\begin{definition}[Cost of Switching]
We define the cost of switching $h$ to $h'$ through $\hat{h}$ as \begin{align*}
c_{switch}(h\rightarrow\hat{h}\rightarrow h') = \log \frac{1}{\Pr[h|\hat{h}]} + \log \frac{1}{\Pr[h'|\hat{h}]}
\end{align*}
The minimal switching cost from $h$ to $h'$ is 
\begin{align*}
c_{switch}(h\rightarrow h') = \min_{\hat{h}\supset h,h'} \log \frac{1}{\Pr[h|\hat{h}]} + \log \frac{1}{\Pr[h'|\hat{h}]}
\end{align*}
\end{definition}

With the above definition, the switching cost is symmetric $c_{switch}(h\rightarrow h')=c_{switch}(h'\rightarrow h)$.

\begin{figure}[h]
    \centering
    \includegraphics[width=0.45\textwidth]{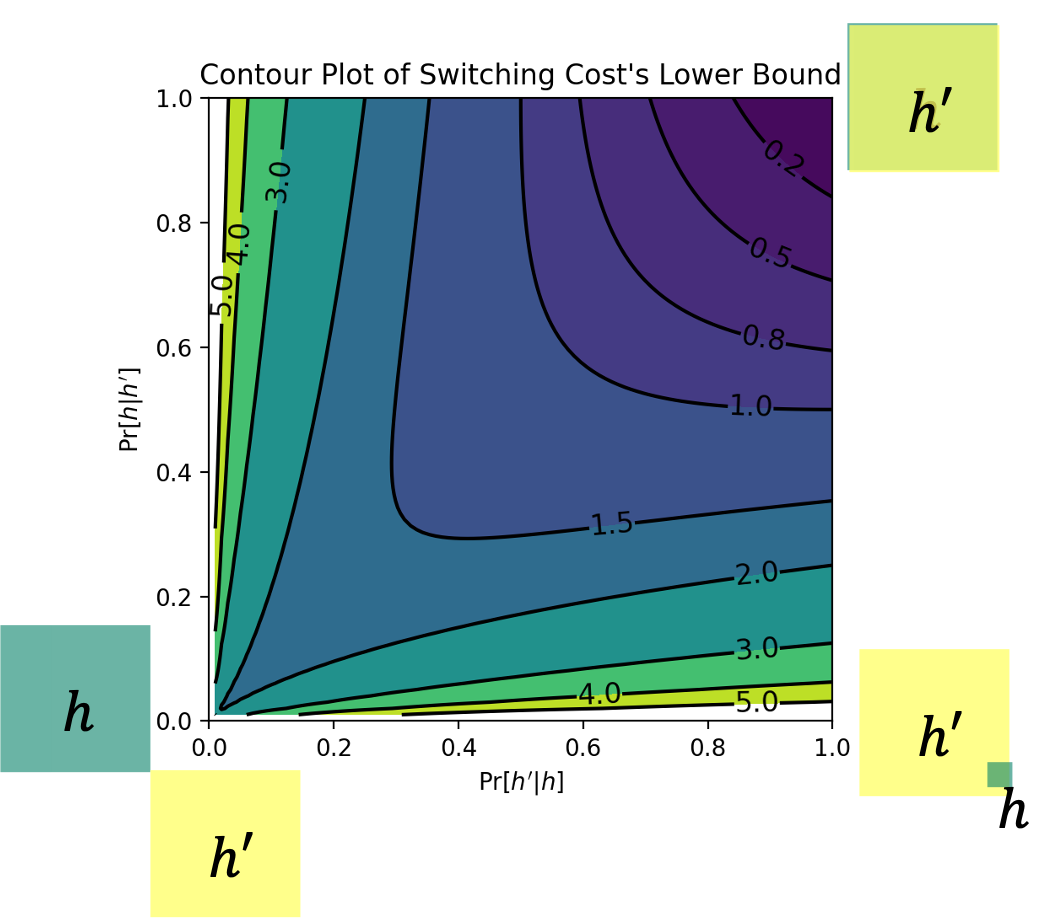} 
     \caption{\textbf{Switching Cost}: From darker to lighter, the lower bound of the switching cost increases. A greater asymmetry, less overlap between two stable states $h$ and $h'$ results in more difficult switching between them. The most extreme case is transitioning from a shallow, superficial understanding to a deep, profound one - the lower right corner. }
    \label{fig:switch} 
\end{figure}

\begin{propositionrep}[Low Switching Cost $\Rightarrow$ Low Disagreement]\label{thm:switch}
Given $\theta,\mathbf{x}$ and two $\epsilon$-consensuses $h,h'$, 
\[c_{switch}(h\rightarrow h')\geq \ell(\Pr[h'|h],\Pr[h|h'])\] where $\ell(x,y)=\log (1+\frac{x}{y}-x)  (1+\frac{y}{x}-y)$. When $c_{switch}(h\rightarrow h')\leq b$ where $b\in[0,1]$, $D_{TV}(\mathbf{q}_h,\mathbf{q}_{h'})\leq \min\{\{2(1-2^{-b}),\sqrt{2^{b+1}g(\theta,\mathbf{x},\epsilon)}\}\}$.
\end{propositionrep}

\begin{appendixproof}

\begin{align*}
c_{switch}(h\rightarrow h') = & \min_{\hat{h}\supset h,h'} \log \frac{1}{\Pr[h|\hat{h}]} + \log \frac{1}{\Pr[h'|\hat{h}]}\\
\geq & \log \frac{1}{\Pr[h|h\cup h']} + \log \frac{1}{\Pr[h'|h\cup h']}\\
= & \log \frac{1}{\frac{\Pr[h]}{\Pr[h]+\Pr[h']-\Pr[h\cap h']}} + \log \frac{1}{\frac{\Pr[h']}{\Pr[h]+\Pr[h']-\Pr[h\cap h']}}\\
= & \log (1+\frac{\Pr[h']}{\Pr[h]}-\Pr[h'|h])  (1+\frac{\Pr[h]}{\Pr[h']}-\Pr[h|h'])\\
= & \log (1+\frac{\Pr[h'|h]}{\Pr[h|h']}-\Pr[h'|h])  (1+\frac{\Pr[h|h']}{\Pr[h'|h]}-\Pr[h|h'])\\
= & \ell(\Pr[h'|h],\Pr[h|h']) \tag{$\ell(x,y)=\log (1+\frac{x}{y}-x)  (1+\frac{y}{x}-y)$}
\end{align*}

We start to analyze the properties of $\ell(x,y)$. We would like to show that when $\ell(x,y)\leq 1$, $\ell(x,y)\geq \max\{ \ell(1,x),\ell(1,y)\},\forall x,y$ where $\ell(1,x)=\log\frac1x,\ell(1,y)=\log\frac1y$. With such a claim, the upper bound $b$ of $\ell(x,y)$ will directly induce a lower bound $2^{-b}$ for both $x,y$. In the current context, $x=\Pr[h'|h], y=\Pr[h|h']$. Therefore, this leads to a lower bound $2^{-b}$ for both $\Pr[h'|h], \Pr[h|h']$. Using \Cref{lem:general}, we obtain the results directly. It's left to prove the claim. 

\begin{claim}
When $\ell(x,y)\leq 1$, $\ell(x,y)\geq \max\{ \ell(1,x),\ell(1,y)\},\forall x,y$.
\end{claim}

Notice that all formulas are algebraic and can be translated into inequalities of low-degree polynomials. We use Mathematica to automatically prove the claim.

\end{appendixproof}

\section{Two More Views of Perception}

This section provides two more views on the communication model - optimization and prediction market. First, communication is cast as an optimization balancing accuracy and information costs. $\beta$-equilibrium corresponds to a local optimum of this goal. Second, an analogy is drawn between perception and markets, viewing markets as communication protocols. This analogy leverages insights from behavioral economics to explain perceptual phenomena involving rivalry and inhibition.

\subsection{An Optimization View: Balancing Accuracy \& Cost}

$\Pi(\mathbf{X})$ is a random variable of final protocol contents on stimulus $\mathbf{x} \sim \mathbf{X}$. $\Pi(\mathbf{X})$ partitions $\mathbf{X}$'s realization space into hyperrectangles (Fig. \ref{fig:partition}). Let $\Pi^t$ be the protocol terminating after $t$ rounds, with final contents $\Pi^t(\mathbf{X})=H^t$. We define a utility measure on each partition. $\Pi^0,\Pi^1,\dots,\Pi^t,\dots,\Pi$ is an optimization process, with final $\Pi$ reaching local optimum at $\beta$-equilibrium. Overall, communication exhibits an optimization dynamic, providing an alternative perspective.

\begin{definition}[Accuracy \& Cost]
Given a communication protocol $\Pi$, we define $\Pi$'s accuracy as the mutual information between $W$ and $\Pi(\mathbf{X})$,
$v(\Pi)=\mi(\Pi(\mathbf{X});W)$;
we define $\Pi$'s information cost as the mutual information between $\Pi(\mathbf{X})$ and $\mathbf{X}$,
$c(\Pi)=\mi(\Pi(\mathbf{X});\mathbf{X})$.
\end{definition}

The definition of $c(\Pi)$ matches External Information Complexity \cite{braverman2013information}, which is closely related to communication complexity \cite{yao1979some}. Literature proves some functions $f(\mathbf{x})$ with $\mathbf{x} \sim \mathbf{X}$ cannot have low complexity protocols with accuracy requirements \cite{kushilevitz1997communication}. This implies for information structure $\theta$ with $W=f(\mathbf{X})$, no $\Pi$ exists with low cost and high value. Intuitively, stimuli drawn from such distributions have hard-to-integrate components. Consequently, the stimulus's meaning is difficult to understand. Overall, this connection shows that communication complexity tools characterize integration difficulty, relating to perceptual difficulty.

To balance accuracy and cost, an optimization goal is \[v(\Pi)-\beta c(\Pi)=\mi(\Pi(\mathbf{X});W)-\beta \mi(\Pi(\mathbf{X});\mathbf{X})\] which can be seen as the application of the information bottleneck framework\footnote{The information bottleneck framework is usually stated as $\min_f \mi(f(\mathbf{X});\mathbf{X}))-\beta \mi(f(\mathbf{X});W)$ where $f$ is compression of inputs $\mathbf{X}$ \cite{tishby2000information}.} into the communication setting.

\paragraph{$\beta$-equilibrium = Local Optimum} We show $\beta$-equilibrium is a local optimum of $v(\Pi)-\beta c(\Pi)$ under proper neighbor definitions. The key observation is that the optimization goal equals the accumulated expected utility of all agents. Each update $\Pi^{t-1}\Rightarrow\Pi^t$ is local. Thus the communication process can be seen as a local optimization algorithm that optimizes total utility, where $\beta$-equilibrium is a local optimum. 

Formally, $h'$ is a neighbor of $h$ if it only differs in one agent's knowledge. $\Pi'$ is a neighbor of $\Pi$ if $\Pi'(\mathbf{x})$ is a neighbor of $\Pi(\mathbf{x})$ $\forall \mathbf{x}$. $\Pi$ reaches $\beta$-equilibrium if $\Pi(\mathbf{x})$ reaches $\beta$-equilibrium $\forall \mathbf{x}$.

\begin{propositionrep}
When $\Pi$ reaches a $\beta$-equilibrium, for all  $\Pi$'s neighbor, $\Pi'$, $v(\Pi')-\beta c(\Pi')\leq v(\Pi)-\beta c(\Pi)$. 
\end{propositionrep}

Information substitutes enable the communication protocol to quickly converge to ground truth belief with minimal cost. In contrast, complements lead to false consensus traps where beliefs get stuck far from the truth. Stimuli based on functions with high external information complexity lead to comprehension difficulties, and full understanding requires escaping poor local optima.

\begin{appendixproof}
We first show the optimization goal $v(\Pi)-\beta c(\Pi)$ is the accumulated expected utility of all agents in the communication. Given $\Pi=(\mathbf{i},\{P_i\}_i)$, $\Pi(\mathbf{X})$ is determined by $\mathbf{X}$. Thus, $c(\Pi)=\en(\Pi(\mathbf{X});\mathbf{X})$. Based on chain rule, $v(\Pi)=\mi(\Pi(\mathbf{X});W)=\mi(H^1;W)+\mi(H^2;W|H^1)+\cdots$. Recall that $H^t=\Pi^t(\mathbf{X})$ are the contents on the board after $t$ rounds. Therefore, $v(\Pi)$ is the accumulated expected reward of all agents in the communication. Analogously, $c(\Pi)=\en(\Pi(\mathbf{X});\mathbf{X})=\en(H^1)+\en(H^2|H^1)+\cdots$. Thus, $\beta c(\Pi)$ is the accumulated expected cost of all agents in the communication. Additionally, $\mi(\mathbf{X};W)-v(\Pi)=\mi(\mathbf{X};W)-\mi(\Pi(\mathbf{X});W)=\mi(\mathbf{X};W|\Pi(\mathbf{X}))$ equals the expected regret of $\Pi(\mathbf{x})$, and $c(\Pi)$ is the expected cost of $\Pi(\mathbf{x})$.

When $\Pi$ reaches a $\beta$-equilibrium, for all $\mathbf{x}$, $\Pi(\mathbf{x})$ reaches a $\beta$-equilibrium. Thus, no agent at $\Pi(\mathbf{x})$ can obtain $>0$ expected utility by revealing her information. Therefore, in average, $v(\Pi')-\beta c(\Pi')\leq v(\Pi)-\beta c(\Pi)$. 
\end{appendixproof}

\subsection{A Prediction Market View} 

A prediction market maintains a state vector $\mathbf{s}$ where $\mathbf{s}(\omega)$ indicates how many $\omega$ shares are in the market. LMSR \cite{hanson2012logarithmic} defines a corresponding relationship between state $\mathbf{s}$ and market price $\mathbf{p}$: $\mathbf{p}(\omega)=\frac{e^{\frac{\mathbf{s}(\omega)}{\alpha}}}{\sum_{z} e^{\frac{\mathbf{s}(z)}{\alpha}}}$ where $\alpha>0$ denotes market liquidity. 

\begin{description}
\item [Trading] At each round $t$, agent $i_t$ moves state from $\mathbf{s}^{t-1}$ to $\mathbf{s}^{t}$. Her cost is $C(\mathbf{s}^{t})-C(\mathbf{s}^{t-1})$ where $C(\mathbf{s})=\alpha \log\left(\sum_z e^{\frac{\mathbf{s}(z)}{\alpha}}\right)$
\item [Reward] When $W=\omega$ is revealed, each $\omega$ share is worth 1 unit price, while others are worth nothing. 
\end{description}

Prior studies \cite{hanson2003combinatorial, hanson2012logarithmic} showed that when $W=\omega$, agent $i_t$'s reward, subtracting the moving cost is 
\[\alpha\left(\log \mathbf{q}_{\mathbf{s}^t}(\omega)-\log \mathbf{q}_{\mathbf{s}^{t-1}}(\omega)\right)\] where $\mathbf{q}_{\mathbf{s}}$ is the market price of state $\mathbf{s}$. Thus, the reward in LMSR matches the reward in the previous communication model. The market does not explicitly model the information cost. Nevertheless, if the traders are not endowed with private information and need to buy the information externally outside the market, then the information cost can be used to model the cost traders need to pay to buy the information. 

Therefore, prediction markets can be seen as a special communication protocol where the blackboard maintains a state vector and the agents can only communicate through modifying the state vector. 

\paragraph{Trader Strategy} Prior studies \cite{chen2010gaming, chen2016informational} showed that with substitute information, traders reveal early to maximize value. Delay causes redundancy that decreases impact. With complementary information, traders delay revelation, as delay increases marginal value over time. 

\paragraph{Impact of Liquidity} One notable feature of prediction markets is liquidity, denoted by $\alpha$ \cite{wolfers2004prediction,arrow2008promise, chen2010new}. Liquidity quantifies the number of shares a trader must trade to meaningfully impact the market price. Markets with low liquidity are sensitive to trading volume - even a small number of shares traded can significantly sway prices. Therefore, markets with low liquidity lead to market instability. In contrast, markets with high liquidity are more resistant to fluctuation. Traders must exchange a substantial volume of shares to noticeably influence prices in a highly liquid market. However, markets with high liquidity may lead to slow convergence. 

\paragraph{Market-Mind Analogy} Analogies include:
\begin{itemize}
\item Market Price - Global Belief: The market price represents a consensus belief for $W$, formed from the cumulative interactions of many traders, just as a mind's beliefs arise from the complex dynamics between neurons. 
\item Market Equilibrium - Stable Perception: When a market lacks significant activity or price fluctuation, market price converges, just as a stable perception emerges. 
\item Traders - Neurons: The trader analogy suggests strategic neural coordination and competition, explaining phenomena like rivalry and inhibition. In binocular rivalry \cite{tong2006neural}, alternating perception between conflicting inputs may arise from neurons suppressing an image, like traders withholding complements. These neurons could strategically time signals to gain advantage when their image dominates. \emph{This is analogous to facing two jigsaw puzzles - a greedy strategy prioritizes completing one puzzle first before starting the other.} Likewise, inhibition \cite{aron2007neural} resembles speculators silencing irrelevant signals when consensus forms. The trader analogy provides a framework for how competitive, strategic neural interactions shape perception - much like markets reaching equilibrium.
\item Market Liquidity - Sensitivity: The liquidity analogy potentially links market dynamics to mental health symptoms. Low liquidity, characterized by volatile belief fluctuations, hypersensitivity to stimuli, and failure of inhibition, mirrors aspects of schizoaffective disorders \cite{frith1979consciousness}. In contrast, high liquidity displays abnormally stable beliefs, and limited responsiveness, similar to patterns observed in autism spectrum disorders that exhibit restricted reactivity to external information \cite{geschwind2007autism}. 
\end{itemize}

Other possible analogies include \emph{Information Cascades - Confirmation Bias} when the initial information almost decides the stable perception because of herding \cite{BHW1992, wason1960failure}, \emph{Market Bubbles - Optimism Bias} when the market price is inflated, analogous to brains skewed toward positive overestimation \cite{smith1988bubbles, sharot2011optimism}.

In addition to the above discussions, prior work also has drawn conceptual connections between markets and mind \cite{schotanus2022cognitive}. The global market behavior is similar to conscious emergence, while local trades are like unconscious processes \cite{sornette2009stock}. Topological analyses reveal isomorphic structures between brain and financial networks \cite{vertes2011topological}. Learning in cortical models shares features with prediction markets \cite{balduzzi2014cortical}. However, it is important to acknowledge that the human brain is complex. While not a perfect metaphor, the market-mind analogy creates an interdisciplinary bridge to compare information processing in markets and the human mind.

\section{Conclusion and Discussion}

This work presented a distributed communication model to investigate multistable perceptions. Framing perception as consensus formation provides tools to analyze how relationships shape collective neural computation toward coherent interpretations. A key finding is substitute versus complementary relationships influence monostable versus multistable perceptions. This work additionally provides an optimization-based and a market-based perspective, which demonstrates how relationships influence perception costs and impact competition behaviors in neural dynamics. Future directions include testing whether the current ``black-box'' machines, including the large language model (LLM) \cite{brown2020language}, display phenomena of multistable perceptions and order effects and designing new machines based on distributed consensus principles to simulate perceptual phenomena. Bridging the distributed communication framework with models of cellular differentiation in evolution may yield new cross-disciplinary insights \cite{waddington2014strategy}. Detectives reaching consensus resemble cells adopting specialized functions to form unified systems during development. Hopefully, this interdisciplinary framework could encourage researchers from diverse fields to unravel the intricacies of perception and cognition together.

\bibliographystyle{plain}
\bibliography{files/ref}
\appendix
\section{Basics of Information Theory}
We introduce multiple basic concepts in information theory. At a high level, entropy gauges intrinsic uncertainty, conditional entropy measures leftover uncertainty, mutual information captures inter-variable correlations, and conditional mutual information assesses residual correlations beyond a third variable.

\begin{definition}[Entropy \cite{shannon1948mathematical}]
The entropy of a random variable $X$ is
$$\en(X):=-\sum_x \Pr[X=x]\log(\Pr[X=x]).$$
The conditional entropy of $X$ conditioning on an additional random variable $Z=z$ is
$$ \en(X|Z=z):=-\sum_x\Pr[X=x|Z=z]\log(\Pr[X=x|Z=z]) $$
The conditional entropy of $X$ conditioning on $Z$ is
$$ \en(X|Z):=\E_{Z}[\en(X|Z=z)]. $$
\end{definition} 

\begin{definition}[Mutual information \cite{shannon1948mathematical}]
The mutual information between two random variables $X$ and $Y$ is
$$\mi (X;Y):=\sum_{X,Y}\Pr[X=x,Y=y]\log\left(\frac{\Pr[X=x,Y=y]}{\Pr[X=x]Pr[Y=y]}\right)$$
The conditional mutual information between random variables $X$ and $Y$ conditioning on an additional random variable $Z=z$ is
$$\mi (X;Y|Z=z):=\sum_{x,y}\Pr[X=x,Y=y|Z=z]\log\left(\frac{\Pr[X=x,Y=y|Z=z]}{\Pr[X=x|Z=z]\Pr[Y=y|Z=z]}\right)$$
The conditional mutual information between $X$ and $Y$ conditioning on $Z$ is
\[\mi(X;Y|Z):=\E_{Z} \mi(X;Y|Z=z).\]
\end{definition}

\begin{fact}[Useful Properties]~\cite{cover2006elements}\label{cor:MI-prop}
\begin{description}
\item[Symmetry:] $\mi(X;Y) = \mi(Y;X)$
\item[Relation to entropy:] $\en(X)-\en(X|Y)=\mi(X;Y)$, $\en(X|Z)-\en(X|Y,Z)=\mi(X;Y|Z)$
\item[Non-negativity:] $\mi(X;Y)\geq 0$
\item[Chain rule:] $\mi(X,Y;Z)=\mi(X;Z)+\mi(Y;Z|X)$
\item[Monotonicity:] when $X$ and $Z$ are independent conditioning on $Y$, $\mi(X,Z)\leq \mi(X;Y)$.
\end{description}
\end{fact}

Symmetry means that mutual information has no directionality. Relation to entropy shows mutual information is the reduction in uncertainty about one variable given the other. Non-negativity means mutual information is zero or positive. Variables cannot have negative mutual information. The chain rule means the mutual information with a pair can be decomposed. Monotonicity means when one side becomes noisy, the two sides' mutual information decreases.

\end{document}